\documentclass[aps, prl,10pt,notitlepage,nofootinbib,superscriptaddress,showkeys,twocolumn,showpacs]{revtex4}

\usepackage[latin1]{inputenc}
\usepackage{fancyhdr}
\usepackage{graphicx}
\usepackage{hyperref}

\usepackage{amssymb,amsmath,mathrsfs,mathtools,amsthm,braket}

\newcommand{\NN}{\mathbb{N}}
\newcommand{\cG}{{\cal G}}

\allowdisplaybreaks[4]

\begin{document}

\title{\large \bf The double scaling limit of the multi-orientable tensor model}

\author{{Razvan Gurau}}\email{rgurau@cpht.polytechnique.fr}
\affiliation{CPHT, UMR CNRS 7644, \'Ecole Polytechnique, Univ. Paris-Saclay, 
91128 Palaiseau cedex, France, EU}
\affiliation{Perimeter Institute for Theoretical Physics, 
31 Caroline st. N, N2L 2Y5, Waterloo, ON, Canada}
\author{{Adrian Tanasa}}\email{adrian.tanasa@lipn.univ-paris13.fr}
\affiliation{LIPN, UMR CNRS 7030, Univ. Paris 13, Sorbonne Paris Cit\'e,
99 av. Cl\'ement, 93430 Villetaneuse, France, EU}
\affiliation{H. Hulubei National Institute for Physics and Nuclear Physics, 
PO Box MG-6, 077125 Magurele, Romania, EU}
\author{{Donald R. Youmans}}\email{donald.youmans@gmx.de}
\affiliation{CPHT, UMR CNRS 7644, \'Ecole Polytechnique, Univ. Paris-Saclay, 
91128 Palaiseau cedex, France, EU}
\affiliation{LIPN, UMR CNRS 7030, Univ. Paris 13, Sorbonne Paris Cit\'e,
99 av. Cl\'ement, 93430 Villetaneuse, France, EU}

\date{\small\today}

\begin{abstract}\noindent
In this paper we study the double scaling limit of the multi-orientable tensor model. We prove that,
contrary to the case of matrix models but similarly to the case of invariant tensor models, the double scaling series 
are convergent. We resum the double scaling series of the two point function and of the leading 
singular part of the four point function. We discuss the behavior of the leading singular part of arbitrary correlation functions.
We show that the contribution of the four point function and of all the higher point functions are enhanced in the double scaling limit. 
We finally show that all the correlation functions exhibit a singularity at the same critical value of the double scaling parameter
which, combined with the convergence of the double scaling series, suggest the existence of a triple scaling limit.
\end{abstract}

\pacs{11.10.Kk,03.70.+k,11.10.Jj}

\maketitle

\section{Introduction}

Tensor models (see \cite{Rivasseau:2013uca,review,Tanasa:2012re} for a detailed introduction) generalize
matrix models \cite{DiFrancesco:1993nw} in dimension higher than two. 
Matrix models posses a $1/N$ expansion \cite{David:1984tx,Kazakov:1985ds} dominated by planar graphs \cite{Brezin:1977sv} and a double scaling 
limit \cite{double,double1,double2}. In this limit one sends $N$ (the matrix size) to infinity and
the coupling constant $\lambda$ to some critical value $\lambda_c$ in such a way that arbitrary genus surfaces 
contribute to a $2r-$point function. This is achieved if one keeps the double scaling parameter $ N^{\frac{5}{4}}(\lambda_c - \lambda)$ fixed. 
As in this regime all the topologies contribute, the double scaling limit 
corresponds to a theory at finite Newton coupling \cite{EDTDavid}. It should be emphasized that, in the case of matrices, 
the double scaling series are \emph{divergent} \cite{DiFrancesco:1993nw}.

Tensor models \cite{review} generalize matrix models in higher dimension. 
They posses a $1/N$ expansion (where $N$ is now the tensor size) \cite{expansion3,expansioin5} and reach a continuum limit \cite{review} by
tuning to criticality. The $1/N$ expansion of tensor models is indexed by the \emph{degree},
which, contrary to the genus, is not a topological invariant.

Tensor models fall into two broad categories: the invariant models \cite{uncoloring} which exist for tensors of arbitrary rank
and the multi-orientable (MO) one \cite{Tanasa:2011ur}. While the MO model exists
only for tensors of rank three, it exhibits a richer combinatorics than the invariant models in rank three: the class of 
graphs the MO model sums over is strictly larger than the one the invariant models sum over. 
In particular, for the multi-orientable model the degree is a \emph{halfinteger}, while it can only be an integer for the invariant models.

The double scaling limit has recently been studied for the invariant tensor models \cite{DGR,Bonzom:2014oua,GurSch}. 
It has been found that, denoting $D\ge 3$ the rank of the tensor, the appropriate double scaling parameter is 
$N^{D-2} (\lambda_c - \lambda)$ for $D<6$ and $N^{\frac{2}{3}D}(\lambda_c - \lambda)$ for $D\ge 6$. The double scaling series are \emph{convergent} for $D<6$ 
and divergent (as they are for matrices) for $D\ge 6$.

In order to identify what features of the double scaling limit of tensor models are robust, 
we study in this paper the double scaling limit of the MO model. 
As this model sums over a larger family of graphs than the invariant models in rank three,
it is the ideal testing ground for this study. We find that, as it was the case for the invariant models in 
rank three, the double scaling series in the MO model are convergent. We perform below the analytic resummation of these series for 
the two point function, the leading singular part of the four point function, and discuss the leading singular part of arbitrary correlation functions. 
Surprisingly, although the convergence of the double scaling series is robust, the double scaling
parameter is \emph{not}. We show that for the MO model the appropriate double scaling parameter is 
$N^{\frac{1}{2}}  (\lambda_c - \lambda) $, hence scales with an exponent of $N$ different from the one of the invariant models in 
rank three.

Finally, we show that all the (leading singular parts of the) correlation functions become critical at the same critical value of the double scaling parameter and
discuss the consequences of this behavior.

\section{The multi-orientable tensor model}

Let $\phi_{ijk}$, $i,j,k = 1, \ldots, N$, be the components of a rank three complex tensor $\phi$. 
The action of the MO tensor model is  \cite{expansioin5}:
\begin{equation}\label{eq:S}
 S[\phi] =  \sum_{ijk} \phi_{ijk} \bar{\phi}_{ijk} - \frac{\lambda}{2N^{3/2}} \sum_{\genfrac{}{}{0pt}{}{ijk}{lmn}}\phi_{ijk} \bar{\phi}_{mlk} \phi_{mjn} \bar{\phi}_{iln} \;,
\end{equation}
where the scaling with $N$ of the interaction term above is chosen 
in order for the model to admit a $1/N$ expansion \cite{expansioin5}.

\begin{figure}[htb]
\includegraphics[scale=0.3]{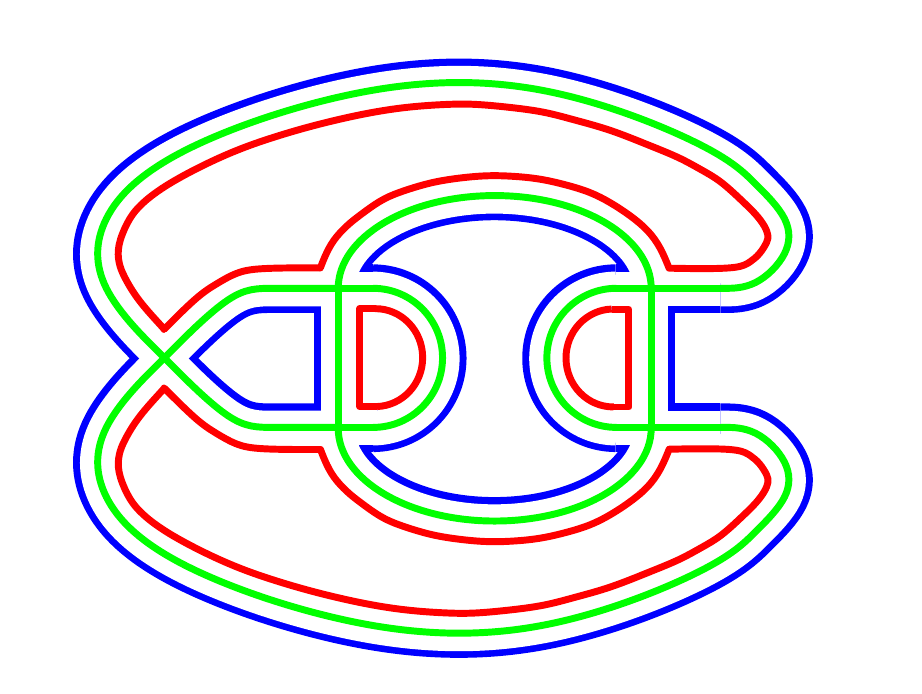}
\caption{\label{graf} An MO vacuum (Feynman) graph.}
\end{figure}

The partition function of the MO model:
\begin{equation*}
 Z = \int \mathcal{D}[\phi] e^{ -S[\phi] } \; , \;\; \displaystyle \mathcal{D}[\phi] = \prod_{ijk} \frac{{\rm d}\phi_{ijk} {\rm ~d}\bar{\phi}_{ijk}}{2\pi\imath} \;,
\end{equation*}
writes in perturbation theory as a sum over Feynman multi-orientable graphs \cite{Tanasa:2011ur}. 
The vertices of the MO graphs are four valent, the edges are oriented from $\phi$ to $\bar \phi$ and  
the orientations alternate around a vertex. The edges can be represented as three parallel strands (one 
for each index of the tensor) and the vertices as the intersection of four half edges such that 
every pair of half edges share a strand. This leads to the representation in Fig.~\ref{graf}. 

The strands are divided into three classes:
the ones in the middle of the edges, called the $S$ (for straight) strands, 
the ones on the right (with respect to the $\phi\to \bar \phi$ orientation) of the edges, called the $R$ (for right) strands
and the ones on the left of the edges, called the $L$ (for left) strands. At a vertex the $S$ ($R$ or $L$) strands
only connect to $S$ ($R$ or $L$) strands.

\begin{figure}[htb]
\includegraphics[scale=0.25]{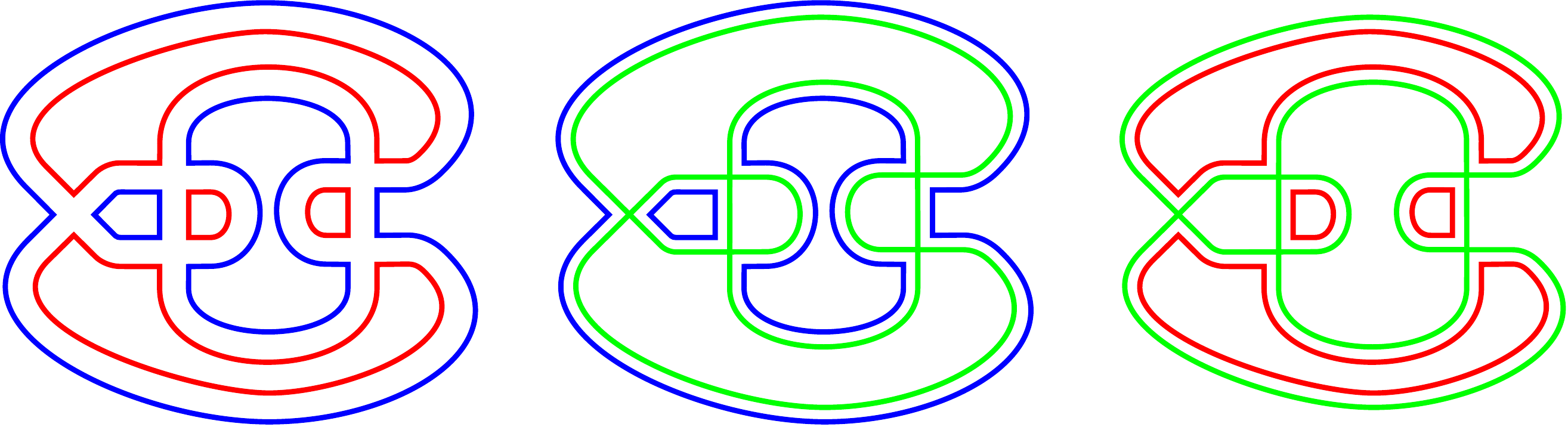}
\caption{\label{jachete}The three jackets associated to the graph of Fig.~\ref{graf}.}
\end{figure}

To any MO graph one can associate three canonical ribbon graphs, called the \emph{jackets},
obtained by erasing throughout the graph all the strands in the same class ($S$, $L$ or $R$). 
For the graph of Fig.~\ref{graf} for example this leads to the three jackets represented in Fig.~\ref{jachete}.
Crucially, the jackets of the MO graphs can be non-orientable (dual to non orientable surfaces).
This is a fundamental difference between the MO model and the invariant models.

\emph{The degree $\delta(\cG)$ of a connected MO graph $\cG$ is the half sum of the non 
orientable genera of its three jackets.} It is therefore a positive half integer.

The Feynman amplitude of the graph $\cG$ is \cite{expansioin5} 
$
\mathcal{A}(\mathcal{G}) = \lambda^{v(\cG)} N^{3-\delta(\cG)},
$
where $v(\cG)$ is the number of vertices of $\cG$. The free energy, which is a sum over connected graphs, admits the $1/N$ expansion:
\begin{equation*}
 \ln Z    = \sum_{\mathcal{G}}  \mathcal{A}(\mathcal{G}) = \sum_{\delta \in \NN / 2} N^{3 - \delta} 
 \Big(  \sum_{\mathcal{G} | \delta(\mathcal{G}) = \delta}  \lambda^{v (\mathcal{G} ) } \Big)   \; .
\end{equation*}
 
In the sequel we will use a simplified representation of the MO graphs (introduced in \cite{Fusy:2014rba}) as graphs with oriented edges.
\section{Correlation functions}
 
Let us denote $ \Braket{\bar \phi \dots \phi \dots}_c$ the connected correlation functions of the MO model.
The action in Eq.~\eqref{eq:S} is invariant under the field transformation:
\begin{align*}
& \phi'_{i_1j_1k_1} = \sum_{ijk}  U^{R}_{i_1i} O^S_{j_1 j} U^L_{k_1k}\phi_{ijk} \;, \crcr
& \bar \phi'_{i_1j_1k_1} = \sum_{ijk} \bar U^{R}_{i_1i} O^S_{j_1 j} \bar U^L_{k_1k} \phi_{ijk}  \;,
\end{align*}
where $U^R$ and $U^L$ are two (distinct) complex unitary matrices and $O^S$ is a real orthogonal matrix. 
The indices $S$ are special: one must use an orthogonal transformation because the quadratic part contracts 
and index $S$ of a field $\phi$ with an index $S$ of a field $\bar \phi$ while the interaction contracts two indices $S$ 
belonging either both to fields $\phi$ or both to fields $\bar \phi$.
 
\begin{figure}[htb]
\includegraphics[scale=0.75]{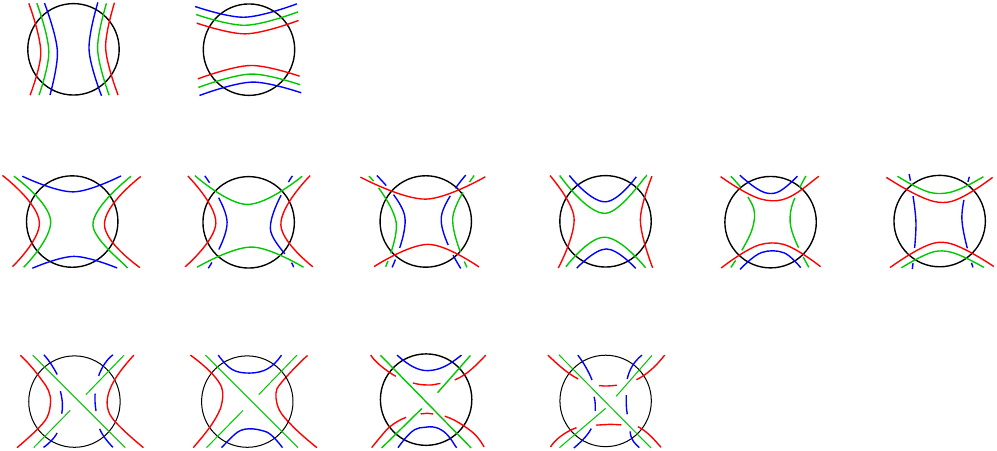}
\caption{\label{fourpoint}The twelve index pairings of the four point function.}
\end{figure}

This invariance constrains the correlation functions: 
averaging over the unitary and orthogonal groups \cite{ColSni} one finds that the indices of any correlation function 
must be pairwise identified. For the $R$ and $L$ indices, an index of a $\phi$ must be identified with an index of a $\bar \phi$,
while for the $S$ indices any pairing is allowed. Thus:
\begin{equation} \label{eq:2pointstrand}
  \Braket{  \bar{\phi}_{\bar{a}\bar{b}\bar{c}}  \phi_ {abc} }_c  = \delta^{\bar{a}}_{a} \delta^{ \bar{b}}_{ b} \delta^{ \bar{c}}_{ c }   \; \mathfrak{K}_2 \;,
\end{equation}
twelve pairings are allowed (see Fig.~\ref{fourpoint}) for a four point function 
and $2^{-r}r! (2r)!$ pairings for a $2r$ point function.
 Out of these pairings, $r!$ correspond to a pairing of the external legs (the top row in 
Fig.~\ref{fourpoint}):
 \begin{align}\label{eq:2rpointstrand}
 & \Braket{ \prod_{i=1}^r  \bar{\phi}_{\bar{a}_i\bar{b}_i\bar{c}_i}  \phi_ {a_ib_ic_i} }_c  = \crcr
 & \qquad = \mathfrak{K}^{(1)}_{2r}   \sum_{\sigma} \prod_{i=1}^r \delta^{\bar a_i}_{a_{\sigma(i)} } \delta^{\bar b_i}_{ b_{\sigma(i)}}  \delta^{\bar c_i}_{  c_{\sigma(i)}}    + \dots \;,
 \end{align}
where $\sigma$ runs over permutations of $r$ elements.
 
The two point function is related to $\ln Z$ by the Schwinger Dyson equation:
\begin{align*} 
& 0 =  \frac{1}{Z}\int \mathcal{D}[\phi] \sum_{abc} \frac{ \partial \left( \phi_{abc} \; e^{-S[\phi]} \right) }{\partial \phi_{abc} }  \Rightarrow \crcr
& \qquad \Rightarrow \mathfrak{K}_2 = 1  + N^{-3} 2\lambda \partial_\lambda \ln Z \;.
\end{align*}
 
This equation is interpreted as follows: $\ln Z$ is a sum over graphs and 
$  \partial_\lambda$ acting on $\ln Z$ marks a vertex. Together with the factor $2$ this comes to 
marking one of the edges incoming at the vertex and $\mathfrak{K}_2 $ is a sum over  MO graphs \emph{rooted} 
at an edge, $ \cG^{\times}$:
\begin{equation}\label{eq:large_N_exp_K_2}
\mathfrak{K}_2 = 1 + \sum_{\delta \in \NN / 2} N^{- \delta} 
 \Big(  \sum_{ \cG^{\times} | \delta( \cG^{\times} ) = \delta}  \lambda^{v (\mathcal{ \cG^{\times}} ) } \Big) \;.
\end{equation}

The constant term $1$ represents the ring graph consisting in the root edge closed onto itself and can be included in the sum.
The ring graph has three faces, one edge and no vertex hence it has degree zero. 

Equation \eqref{eq:large_N_exp_K_2} can also be understood as follows. 
Let us denote $(\bar \phi   \phi)$ the invariant $\sum_{abc} \bar \phi_{abc} \phi_{abc}$.
The expectation $\braket{ (\bar \phi   \phi)  }_c  $ is precisely a 
sum over connected MO graphs having a special edge corresponding to the insertion $(\bar \phi   \phi) $  which is taken as root. 
Eq.~\eqref{eq:2pointstrand} leads to $ N^3 \mathfrak{K}_2 = \braket{ (\bar \phi   \phi)  }_c $ and Eq.~\eqref{eq:large_N_exp_K_2} follows.  

Similarly, $\Braket{ (\bar \phi \phi)^r }_c$ is a sum over graphs with $r$ root edges and 
Eq.~\eqref{eq:2rpointstrand} leads to:
\begin{equation*}
 \Braket{ (\bar \phi \phi)^r }_c =  \mathfrak{K}^{(1)}_{2r}  \Big( N^{3r} + O(N^{3r-1})  \Big) + \dots \; .
\end{equation*}

 \section{Classification by degree} 
\begin{figure}[htb]
\includegraphics[scale=.6]{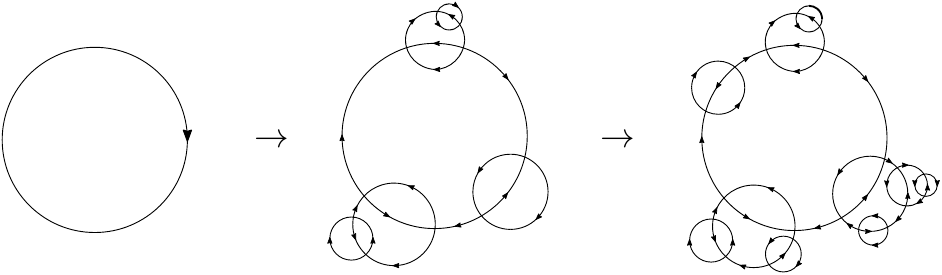}
\caption{\label{fig:melonic}Melonic graphs.}
\end{figure}

We first discuss graphs with a unique root edge.
Let us call a \emph{fundamental submelon} of $\cG^{\times}$ a subgraph consisting in two vertices 
connected by three non-root edges such that the two vertices belong to \emph{three} faces of length two.
The degree of $\cG^{\times}$ does not change if one replaces a fundamental submelon by an edge.
The \emph{melonic graphs} \cite{critical} are
the graphs which reduce to the ring graph by eliminating iteratively all the fundamental submelons (see Fig.~\ref{fig:melonic}) 
and therefore have degree zero.

It turns out that the converse is also true \cite{critical}: 
a graph has degree zero if and only if is it is melonic. The generating function of rooted melonic graphs with respect to the number of vertices, $ T(\lambda^2) $, 
obeys the equation:
\begin{equation*}
  T(\lambda^2) = 1 + \lambda^2 [ T(\lambda^2) ]^{4} \;.
\end{equation*}
 
\paragraph{Cores} 
Deleting iteratively all the fundamental submelons a graph $\cG^{\times}$ reduces to its \emph{core} $\hat \cG^{\times}$. 
The core of the melonic graphs is the ring graph. The set of all the cores consists in all the MO graphs having no fundamental submelon.  

Of course there are many graphs corresponding to the same core and 
the sum over graphs can be reorganized by cores:
$
 \sum_{\cG^{\times}} \lambda^{v(\cG^{\times})}= \sum_{\hat \cG^{\times}} \left(\sum_{\cG^{\times} \text{ with core } \hat \cG^{\times}}  \lambda^{v(\cG^{\times})}\right) 
$. The graphs corresponding to a core are obtained by inserting melonic graphs on the edges of the core:
\begin{equation*}
    \sum_{\cG^{\times} \text{ with core } \hat \cG^{\times}}  \lambda^{v(\cG^{\times})} = \lambda^{v(\hat \cG^{\times})}  \left[ T(\lambda^2) \right]^{2v(\hat \cG^{\times})+1} \;,
\end{equation*}
where we accounted for insertions of melonic graphs at both ends of the root edge of the core.

\paragraph{Schemes}
The classification of graphs in terms of cores is not sufficient because there are still an infinite number of cores of the same degree.

\begin{figure}[htb]
\includegraphics[scale=.5]{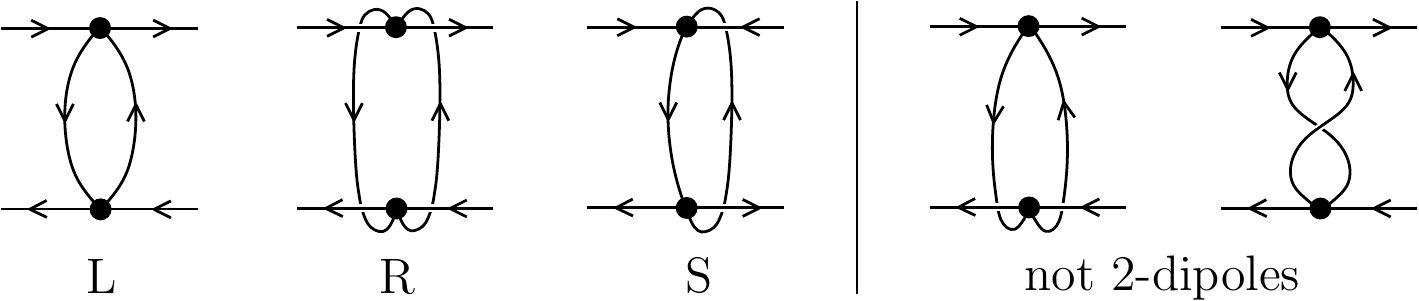}
\caption{\label{dipoli}The three types of dipoles, on the left. On the right, subgraphs which are not dipoles.}
\end{figure}

Let us call a \emph{dipole} a sub graph made of two vertices connected by two non-root edges 
such that the two vertices belong to \emph{exactly one} face of length two. As the face of length two can be of type $L$, $R$ or $S$, there are
three types of dipoles, depicted in Fig.~\ref{dipoli}.
The dipoles can join together into \emph{chains} of dipoles.
A chain is called \emph{broken} (and denoted $B$) if the dipoles in the chain are not of the same
type and \emph{unbroken} if they are. The unbroken chains can be $R$, $L$, or $S$ and the unbroken chains of type $S$ are further 
classified into those with an odd number of dipoles, $S_o$, and those with an even number of dipoles, $S_e$. 

Replacing a chain of dipoles by a chain of the same type but having a different number of dipoles 
does not change the degree of a graph. This is why the number of cores of the same degree is infinite.
In order to classify the graphs with the same degree into only a finite number of classes
one needs to gather in the same class all the cores which only differ by the length of the 
chains of dipoles. Such a class is represented by a \emph{scheme}, which is a rooted MO graph with no fundamental submelons and no chains of 
dipoles, but having, on top of the usual $\lambda$ vertices, five kinds of \emph{chain vertices} representing the five kinds of chains: $B,R,L,S_o,S_e$. 

Let us denote $U(\lambda^2) = \lambda^2 [ T(\lambda^2) ]^{4} $ the generating function of nonempty melonic graphs. We denote $v$ 
the total number of usual vertices and $c$, $b$, $s$ and $s_o$ the total numbers of chain vertices, broken chain vertices, straight chain vertices and straight odd chain vertices in a scheme. 
The generating function of rooted MO graphs corresponding to the scheme $S$ is \cite{Fusy:2014rba}:
\begin{align*}
 F_S^{(\delta)}(\lambda) =   \frac{  T(\lambda^2)  6^{b } [ U(\lambda^2)]^{\frac{v}{2} + 2 c + s_o } }{ [1-U(\lambda^2) ]^{c-s }[1- U(\lambda^2)^2 ]^{s}[1-3U(\lambda^2)]^{b }} \;. 
\end{align*}

Graphs with $r$ root edges are dealt with identically. They are classified by schemes with $r$ roots and the generating function of graphs with $r$ roots
corresponding to a scheme with $r$ roots is:
\begin{align*}
 F_S^{(\delta),r}(\lambda) =   \frac{  T(\lambda^2)^r  6^{b } [ U(\lambda^2)]^{\frac{v}{2} + 2 c + s_o } }{ [1-U(\lambda^2) ]^{c-s }[1- U(\lambda^2)^2 ]^{s}[1-3U(\lambda^2)]^{b }} \;. 
\end{align*}

The main gain in classifying MO graphs by schemes is that the set ${\cal S}_{\delta,r}$ of the schemes of degree $\delta$ with $r$ roots is \emph{finite} \cite{Fusy:2014rba}.

\section{Dominant schemes} The generating function of rooted melonic graphs becomes critical for $\lambda^2_c = 3^3/2^8$. 
In this regime:
\begin{align*} 
 T(\lambda^2)  &\sim \frac{1}{3} \left( 4 - \sqrt{\frac{8}{3} } \sqrt{1 - \frac{\lambda^2}{\lambda_c^2}} \right) \; ,\crcr
 1 - 3U(\lambda^2) & \sim \sqrt{\frac{8}{3}} \left(1 - \frac{\lambda^2}{ \lambda_c^2} \right)^{1/2} \; .
\end{align*}

The generating functions of graphs corresponding to a scheme $F_S^{(\delta),r}(\lambda)$ (and consequently the ones of graphs of fixed degree) become
singular. The leading singular behavior is given by the schemes which maximize the number of broken chains at fixed degree $\delta$,
which we call \emph{dominant}. 

The \emph{deletion} of a chain vertex consists in deleting the chain vertex and joining the two pairs 
of half edges at the same end of the chain vertex  into two new edges. A chain vertex is called \emph{separating} if by 
deleting it the scheme separates into two connected components and \emph{non separating} if not. 

It can be shown \cite{Fusy:2014rba} that by deleting a 
non separating chain vertex the degree of a scheme strictly decreases, while by deleting a separating chain vertex 
the degree of the scheme is distributed among the connected components.
It follows that all the chain vertices in a
dominant scheme must be separating, and (so as to maximize the number of broken ones) all of them must be broken.

The dominant schemes have then the structure of an abstract tree whose edges represent the broken chain vertices 
and whose vertices represent the MO graphs obtained by deleting simultaneously all the broken chain vertices.
The tree is binary so as to maximize the number of broken chain vertices

The three valent nodes of the tree represent MO graphs of degree zero (otherwise one can build a binary tree representing a scheme with 
strictly smaller degree by replacing such a node by a MO graph of degree zero). There are four possible graphs: either the ring graph or 
three melonic graphs with two vertices. 

The univalent nodes of the tree can be of two types: they either represent a ring graph consisting in a root edge
or they represent a MO graph with exactly one vertex and two edges (a ``double tadpole graph''\cite{Fusy:2014rba}) which has degree $1/2$. There are two 
possible double tadpole graphs. The structure of a dominant scheme with several roots is then presented in Fig.~\ref{dominantscheme} below.
\begin{figure}[htb]
\includegraphics[scale=.7]{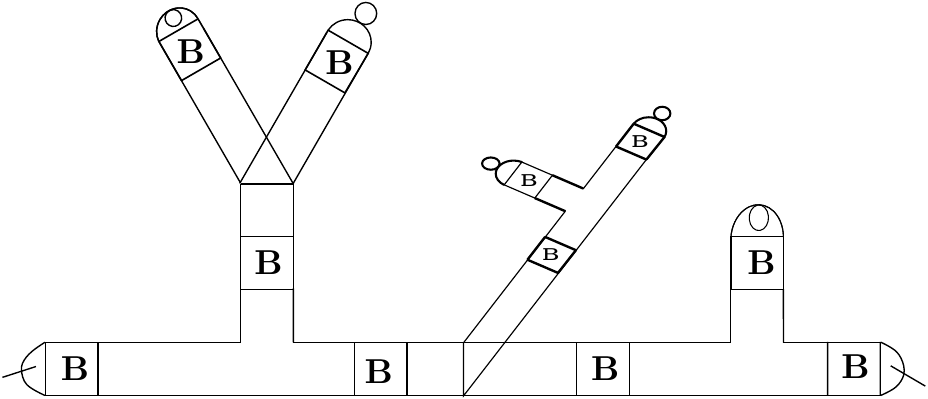}
\caption{\label{dominantscheme}A dominant scheme}
\end{figure}

The dominant schemes with $r$ root edges contribute to the $2r-$point function. As all the chain vertices in such a scheme are broken, 
the dominant schemes contributed only to $ \mathfrak{K}^{(1)}_{2r}$.

The set of dominant schemes with degree $\delta$ and $r$ roots, $ {\cal S}^{\text{dom}}_{\delta, r}$, 
consists in binary trees with $r$ univalent root vertices and another $2\delta$ univalent vertices. Such trees have $2\delta + r -2$
three valent internal vertices and $ 4\delta + 2r - 3$ edges. The leading singular contribution
to the $2r-$point function is then:
\begin{align}\label{eq:domi}
&  \mathfrak{K}^{(1)\text{sing}}_{2r} = N^{3(1-r)} \sum_{\delta \in \NN / 2}   N^{ - \delta }   
\sum_{S \in {\cal S}^{\text{dom}}_{\delta, r} } \crcr 
& \qquad T(\lambda^2)^r [2^2   U(\lambda^2)  ]^{\delta} [1+3 U(\lambda^2) ]^{2\delta+r-2}  \times \crcr
& \qquad \times \left(  \frac{ 
   6  [ U(\lambda^2)]^{ 2 } } { [1-U(\lambda^2) ] [1-3U(\lambda^2)] } \right)^{4\delta + 2r-3} \; ,
\end{align}
where for the two point function ($r=1$) one needs to add the contribution $ T(\lambda^2)$ of the degenerate dominant scheme consisting
in a unique root vertex.

\section{The double scaling limit}

In the double scaling limit one compensates the $1/N$ suppression of the higher order terms in the series in Eq.~\eqref{eq:domi}
by the enhancement at criticality due to the $  [1-3U(\lambda^2)]^{-1} $ factors. This is achieved by sending at the same time $N$ to infinity and
$\lambda$ to criticality while keeping the double scaling parameter
\begin{align*}
N^{\frac{1}{2}} \left( 1 - \frac{\lambda^2}{\lambda_c^2}\right) \equiv \kappa^{-1} \;,
\end{align*}
fixed. In the double scaling regime we get:
\begin{equation*} 
T(\lambda^2) \sim  \frac{4}{3} \left( 1 - \sqrt{\frac{1}{6\kappa\sqrt{N}} }\right) \; , 
1 - 3U(\lambda^2) \sim \sqrt{\frac{8}{3\kappa\sqrt{N}}}.
\end{equation*}

\paragraph{The two point function}

The dominant schemes are rooted binary trees which are counted by the Catalan numbers: there are ${\rm Cat}_{n-1} $ such trees with $n-1$ three valent vertices,
hence with $n$ non root univalent vertices. The degree of the scheme is $ \delta = n/2$ and taking into account the contribution of
the degenerate scheme consisting in a unique root vertex we obtain:
\begin{align*}
&  \mathfrak{K}^{\text{sing}}_{2} = T(\lambda^2) + \sum_{n\ge 1} {\rm Cat}_{n-1} \frac{1}{N^{ \frac{n}{2} } } \times \crcr
 &  \qquad \times  T(\lambda^2)   [2^2   U(\lambda^2)  ]^{\frac{n}{2}} [1+3 U(\lambda^2) ]^{n-1}   \crcr
& \qquad \times \left(  \frac{ 
   6  [ U(\lambda^2)]^{ 2 } } { [1-U(\lambda^2) ] [1-3U(\lambda^2)] } \right)^{2n -1 }   \crcr
& = T(\lambda^2) \left( 1 + \frac{1}{N^{1/2}}\frac{12\cdot U(\lambda^2)^{5/2} }{ [1-U(\lambda^2)][1-3U(\lambda^2)] } \times \right.  \crcr
& \;\; \times \left.\sum_{n\geq 0} {\rm Cat}_n   
\left( \frac{1}{N^{1/2}}\frac{ 72\cdot U(\lambda^2)^{9/2} [1+3U(\lambda^2)]  }{ [ 1-U(\lambda^2) ]^2 [ 1-3U(\lambda)]^2}  \right)^n \right) \;,
\end{align*}
which in the double scaling limit becomes:
\begin{align*}
 \mathfrak{K}^{DS}_2 & = \frac{4}{3}\left( 1 - \sqrt{\frac{1}{6\kappa\sqrt{N}} }\right)  + \crcr
 & \qquad + \frac{4}{3} \frac{1}{N^{\frac{1}{4}} } \sqrt{ \frac{\kappa}{2} } \sum_{n\geq 0} {\rm Cat}_n \left( \kappa \frac{\sqrt{3}}{2}\right)^n    \crcr
 & = \frac{4}{3} \left( 1 - \frac{  \sqrt{1-2\sqrt{3} \kappa} }{ N^{\frac{1}{4} } \sqrt{6 \kappa} }\right) \;,
\end{align*}
where the sum over $n$ converges for $\kappa<2^{-1} 3^{-\frac{1}{2}}$.

\paragraph{The four point function}

The dominant schemes are binary trees with two roots. There are again ${\rm Cat}_{n-1} $ such trees with $n-1$ three valent vertices,
but this time they have $n-1$ non root univalent vertices and degree $\delta = \frac{n-1}{2}$. 
We obtain:
\begin{align*}
& \mathfrak{K}^{(1)\text{sing}}_{4} =  N^{-3}  \sum_{n\ge 1} {\rm Cat}_{n-1} \frac{1}{N^{ \frac{n-1}{2} } } \crcr 
& \quad T(\lambda^2)^2 [2^2   U(\lambda^2)  ]^{\frac{n-1}{2}} [1+3 U(\lambda^2) ]^{n-1}  \times \crcr
& \quad \times \left(  \frac{ 
   6  [ U(\lambda^2)]^{ 2 } } { [1-U(\lambda^2) ] [1-3U(\lambda^2)] } \right)^{2(n-1) + 1} \crcr
   =&  N^{-3}  T(\lambda^2)^2 \left(  \frac{ 
   6  [ U(\lambda^2)]^{ 2 } } { [1-U(\lambda^2) ] [1-3U(\lambda^2)] } \right) \times\crcr
  &  \sum_{n\geq 0} {\rm Cat}_n   
\left( \frac{1}{N^{1/2}}\frac{ 72\cdot U(\lambda^2)^{9/2} [1+3U(\lambda^2)]  }{ [ 1-U(\lambda^2) ]^2 [ 1-3U(\lambda)]^2}  \right)^n \; .
\end{align*}

In the double scaling limit this becomes
\begin{align*}
 \mathfrak{K}^{(1)DS}_{4} = N^{-3 + \frac{1}{4} } \sqrt{\kappa} \frac{8 \sqrt{3}}{ 9\sqrt{2} } \left(  \frac{1 - \sqrt{1-2\sqrt{3}\kappa} }{  \sqrt{3} \kappa } \right) \;.
\end{align*}

Observe that $\mathfrak{K}^{(1)DS}_{4}$ is \emph{enhanced} by a factor $N^{\frac{1}{4}}$ with respect to the natural $N$ scaling
of $\mathfrak{K}^{(1)}_{4} $. This is a consequence of the fact that the singularity of 
the generating function of dominant schemes boosts this four point function in double scaling.

\paragraph{The $2r-$point function}

The dominant schemes are binary trees with $r$ roots and the double scaling limit of $\mathfrak{K}^{(1)}_{2r}$ is:
\begin{align*}
 \mathfrak{K}^{(1)DS}_{2r} \sim & N^{3(1-r)}  N^{\frac{1}{4}(2r-3)} f_{2r}(\kappa) \;,
\end{align*}
for some function $f_{2r}$ depending only on the double scaling parameter $\kappa$. As it was the case for the four point function, the higher point functions
are also boosted in double scaling with respect to their natural scaling in $N$. All the functions $f_{2r}(\kappa) $ are convergent for 
$2\sqrt{3}\kappa < 1 $ and exhibit a square root singularity at the critical double scaling coupling $ \kappa_c= 2^{-1} 3^{-\frac{1}{2}}  $. 

\section{Towards a triple scaling limit}

The results of the previous section open up the possibility to explore a \emph{triple scaling}
regime in which one sends $N \to \infty$, $\lambda \to \lambda_c$ such that $\kappa= \left[  N^{\frac{1}{2}} \left( 1 - \frac{\lambda^2}{\lambda_c^2}\right) \right]^{-1}$,
instead of being fixed, also goes to criticality $\kappa \to \kappa_c = 2^{-1} 3^{-1/2}$ such that 
the triple scaling constant $ N^{\alpha} (\kappa - \kappa_c) = x $, with an appropriate choice of $\alpha$, is kept fixed. 
In this regime the double scaled two point function itself becomes critical and, for a suitable triple scaling exponent $\alpha$, one expects 
an even larger family of graphs to contribute.

\bibliography{Refs}{}

\end{document}